\documentclass[aps,prd,twocolumn,floatfix]{revtex4}
\usepackage{graphicx}
\newcommand{\be}{\begin{equation}}
\newcommand{\ee}{\end{equation}}
\newcommand{\bea}{\begin{eqnarray}}
\newcommand{\eea}{\end{eqnarray}}
\newcommand{\bwt}{\begin{widetext}}
\newcommand{\ewt}{\end{widetext}}
\begin{document}
\title{The Standard Model Higgs Boson-Inflaton and Dark Matter}
\author{T.E. Clark}
\email[e-mail address:]{clark@physics.purdue.edu}
\affiliation{Department of Physics,\\
 Purdue University,\\
 West Lafayette, IN 47907-2036, U.S.A.}
\author{Boyang Liu}
\email[e-mail address:]{liu115@physics.purdue.edu}
\affiliation{Department of Physics,\\
 Purdue University,\\
 West Lafayette, IN 47907-2036, U.S.A.}
\author{S.T. Love}
\email[e-mail address:]{love@physics.purdue.edu}
\affiliation{Department of Physics,\\
 Purdue University,\\
 West Lafayette, IN 47907-2036, U.S.A.}
\author{T. ter Veldhuis}
\email[e-mail address:]{terveldhuis@macalester.edu}
\affiliation{Department of Physics \& Astronomy,\\
 Macalester College,\\
 Saint Paul, MN 55105-1899, U.S.A.}

\begin{abstract}
The standard model  Higgs boson can serve as the inflaton field of slow roll inflationary models provided it exhibits a large non-minimal coupling with the gravitational scalar curvature. The Higgs boson self interactions and its couplings with a standard model singlet scalar serving as the source of dark matter are then subject to cosmological constraints. These bounds, which can be more stringent than those arising from vacuum stability and perturbative triviality alone, still allow values for the Higgs boson mass which should be accessible at the LHC.  As the Higgs boson coupling to the dark matter strengthens, lower values of the Higgs boson mass consistent with the cosmological data are allowed.
\end{abstract}
\maketitle

\section{Introduction}

The inflationary paradigm has proven extremely successful in accounting for a wealth of cosmological observations. It not only explains the large scale homogeneity and isotropy of the present universe, but also accounts for the generation of the nearly scale invariant primordial perturbations responsible for structure formation \cite{inflation}. The implementation of the  slow roll inflation  is  generally achieved by the inclusion of an independent scalar degree of freedom, the inflaton. It is tempting to try to identify the inflaton with the standard model Higgs field, $h$, with a $\lambda h^4$ potential.  Unfortunately, early attempts at such an identification were plagued \cite{Linde,Salopek:1988qh} by the required flatness of the inflaton potential which is dictated by the size of the observed density fluctuations. This in turn necessitated far too small a quartic self coupling for the inflaton for it to be identified with the Higgs scalar. However, by including a  non-minimal coupling of the Higgs doublet, $H$, to the gravitational Ricci scalar curvature, $R$, with a large coupling constant $\xi \sim 10^3 -10^4$, it has been argued \cite{Salopek:1988qh,Bezrukov:2007ep,Barvinsky:2008ia,DeSimone:2008ei,Bezrukov:2008ej,Barvinsky:2009fy} that the resultant Higgs potential can indeed be flat enough for it to play the additional role of the inflaton. In such a case,  the shape of the Higgs-inflaton effective potential in the inflationary region allows for a range of cosmologically acceptable values for the Higgs boson mass even after the inclusion of radiative corrections. As such, slow roll inflation can be implemented without the need for additional degrees of freedom already appearing in the minimal standard model.

The presence of non-baryonic dark matter, however, does require the inclusion of additional degrees of freedom beyond those appearing in the standard model. A minimal extension of the standard model which accounts for this dark matter is the inclusion of a singlet hermitian scalar field, $S$, \cite{Silveira:1985rk,McDonald:1993ex,Burgess:2000yq,Davoudiasl:2004be}. This simple, and clearly minimal, extension can account for the correct abundance of dark matter. Its lack of direct \cite{He:2008qm} and indirect  \cite{Ponton:2008zv,Yaguna:2008hd} detection or observation at high energy accelerators \cite{Bird:2004ts,Barger:2007im,He:2007tt} have already somewhat constrained the model's parameter space, and experiments in the near future will further probe its validity.

In this paper, we study the quantum effects of dark matter as modeled by this standard model singlet scalar on the Higgs-inflaton effective potential. As a result of the coupling of the Higgs boson and dark matter singlet, the cosmological restrictions which include up to three standard deviations from the measured central values, allow for lower values of the Higgs boson mass than is the case in the absence of such dark matter couplings. In section \ref{section2}, the renormalization group improved effective potential for the physical Higgs field is determined including the effects of its coupling to the scalar singlet dark matter as well as its large non-minimal coupling  to the gravitational Ricci scalar.  Three quantities, $\epsilon$, $\eta$ and $\zeta^2$, parameterizing slow roll inflation are given in terms of the derivatives of this Higgs-inflaton effective potential. The Higgs non-minimal coupling $\xi$ to gravity is specified at the onset of inflation so as to yield the measured value for the amplitude of density perturbations.  The scale for the onset of slow roll inflation is set so as to provide $N_e =60$ e-folds of expansion before the inflaton exits inflation at $\epsilon =1$.  The spectral index, its running and the tensor to scalar ratio are then determined quantities which depend on the Higgs boson mass as well as the Higgs boson-dark matter coupling and the singlet self-coupling.  The numerical results for this dependence are presented in section \ref{section3} and are displayed in Figs. \ref{NoRollConstraints} and \ref{CosmoParameterSpace-00} in terms of cosmologically allowed and excluded regions of parameter space.   Conclusions are summarized in section \ref{section4}.  Appendix \ref{appendixA} provides the one-loop renormalization group $\beta$-functions for the gravitationally non-minimally coupled model.  Appendix \ref{appendixB} contains the triviality and vacuum stability constraints for the scalar coupling constants.

\section{Higgs-Inflaton effective potential \label{section2}}

A minimal way of accounting for the  presence of non-baryonic dark matter is by modeling said matter by a stable hermitian standard model singlet, scalar field $S$. The stability of this scalar is guaranteed through the imposition of an unbroken $Z_2$ discrete symmetry under which the scalar is odd.  Focusing on the scalar and gravitational sector of this model the tree level action in the Jordan frame including all terms through mass dimension four is specified by
\bwt
\bea
\Gamma_{\rm Tree} &=& \int d^4x \sqrt{-g}\left[\Lambda +\frac{1}{2}m_{Pl}^2 R +(D_\mu H)^\dagger g^{\mu\nu}D_\nu H -V(H^\dagger H)+\xi R (H^\dagger H -\frac{v^2}{2})\right. \cr
 & &\left. \qquad +\frac{1}{2}\partial_\mu S g^{\mu\nu}\partial_\nu S -\frac{1}{2}m_S^2 S^2 -\frac{\lambda_S}{4}S^4 + \frac{1}{2}\xi_S R S^2 -\kappa S^2 (H^\dagger H -\frac{v^2}{2})+\cdots\right]
\label{treeLag}
\eea
\ewt
where the Higgs potential is
\be
V(H^\dagger H) =\lambda \biggr(H^\dagger H -\frac{v^2}{2}\biggr)^2
\ee
with the Higgs multiplet parameterized by 
\be
H=\frac{1}{\sqrt{2}}\pmatrix{\chi^+ \cr
(v+h ) + \chi^0}  .
\ee
Here $\chi^\pm$, $\chi^0$ are the standard model erstwhile Nambu-Goldstone bosons and $h$ the physical Higgs boson field.  The ellipses in Eq. (\ref{treeLag}) refer to the usual remaining standard model Yang-Mills and fermion  kinetic terms and Yukawa couplings.  Note that this action includes the non-minimal coupling of the Higgs doublet to the gravitational Ricci scalar $R$ with coupling $\xi$.  $m_{Pl}^2$ is the reduced Planck mass $m_{Pl}^2 = 1/8\pi G \approx (2.43 \times 10^{18})^2 ~{\rm GeV}^2$ and $v=246.2$ GeV is the scale of electroweak symmetry breaking.

The various cosmological parameters governing the slow roll inflation are secured in terms of derivatives of the effective potential with respect to the Higgs-inflaton field  evaluated during the inflationary phase when (only) the physical Higgs field has a large expectation value, $h \sim m_{Pl}/\sqrt{\xi}>>v$.  Thus it is necessary to determine the renormalization group improved effective potential as a function of $h$.  The non-minimal gravitational interaction of the Higgs doublet couples  the gravitational and Higgs fields' equations of motion.  This results in a modified form for the high energy behavior of the Higgs and gravitational field propagators or, in the external gravitational field case focused on here, a modified Higgs propagator \cite{Barvinsky:2009fy}.  At the tree level, the non-minimal Einstein equations obtained from (\ref{treeLag}) are
\be
\left( 1+\frac{\xi h^2}{m_{Pl}^2} \right)\left( R_{\mu\nu} -\frac{1}{2} R g_{\mu\nu} \right) = -\frac{1}{m_{Pl}^2}(\Theta_{\mu\nu} + S_{\mu\nu}) ,
\label{Einstein}
\ee
where the new, improved energy-momentum tensor is $\Theta_{\mu\nu} = T_{\mu\nu} + \xi (\frac{1}{2}(\nabla_\mu \nabla_\nu +\nabla_\nu \nabla_\mu) - g_{\mu\nu} \nabla^2 ) h^2$ with $T_{\mu\nu}$ the standard model canonical energy-momentum tensor.  $S_{\mu\nu}$ are terms involving the dark matter field $S$ and its coupling to itself, the gravitational field and the standard model Higgs field.  Focusing on the gravitational and Higgs fields only as they are dominant in the inflationary region, the trace of Eq. (\ref{Einstein}) results in 
\bea
\left( 1+\frac{\xi h^2}{m_{Pl}^2} \right)R &=& \frac{-1}{m_{Pl}^2}\biggr( 6\xi h \nabla^2 h \cr
 & &\qquad + 6(\xi + \frac{1}{6}) \nabla h \cdot \nabla h -4 V(h)\biggr).\cr
 & & 
\eea
Substituting this into the Higgs field equation yields
\bea
\nabla^2 h + V^\prime (h) &=& \xi h R \cr
 &=& \frac{-6\xi^2 h^2 /m_{Pl}^2}{(1 + \xi h^2 /m_{Pl}^2)} \nabla^2 h \cr
 & &\quad -\frac{\xi h ( 6(\xi + 1/6)\nabla h\cdot\nabla h -4V(h)}{(m_{Pl}^2 +\xi h^2)} .\cr
 & & 
\eea
Expanding about the backround Higgs field $h \rightarrow h + \rho$, produces the field equation
\be
\frac{1}{s(h)} \nabla^2 \rho = -V^\prime +\cdots ,
\ee
where (primes denote differentiation with respect to $h$)
\be
s(h)=\frac{1 + \xi h^2 /m_{Pl}^2}{1 +(1+6\xi)\xi h^2 /m_{Pl}^2} .
\ee

Thus the non-minimal coupling to the gravitational field introduces a modification to the Higgs field propagator  by a factor of $s(h)$. The one-loop renormalization group improved effective potential for $h$ can now be computed using standard techniques. The only modification to the standard model calculation arises from the non-canonical Higgs kinetic term. Thus for large non-minimal gravitational coupling constants, $\xi$, and for the sizeable, but still sub-Planckian, inflationary backgrounds $h \sim m_{Pl}/\sqrt{\xi}$, this leads to a suppression of the physical Higgs field propagator \cite{Salopek:1988qh,DeSimone:2008ei,Barvinsky:2009fy}
\be
i/p^2 \rightarrow is/p^2 .
\ee
Note that $s(h )$ takes its canonical value $s(h)\rightarrow 1$ when $\frac{\xi h^2}{m_{Pl}^2}<<1$ while in the other extreme,  
$ s(h)\rightarrow \frac{1}{6\xi}$ when $\frac{\xi h^2}{m_{Pl}^2}>>1$.

It then follows, for these large field values, the Higgs field is no longer an unconstrained high energy degree of freedom. This will result in a violation of unitarity and an (as yet unknown) ultraviolet completion of the model will be necessary for a thorough understanding of the dynamics in the inflationary region \cite{Burgess:2009ea}. Moreover, although $h$ is sub-Plankian in the inflationary region, the variable $\sqrt{\xi} h /m_{Pl}$ is large and the issue of contributions from higher dimensional operators needs further scrutiny. As discussed in reference \cite{Bezrukov:2008ej}, the Higgs suppression could indicate the onset of a new strongly interacting phase of the Higgs sector of the standard model. To account for this, the perturbative low energy standard model renormalization group running of the coupling constants was matched onto their running in a non-linearly realized electroweak symmetry model effective in the range $m_{Pl}/\xi {< }h  {< } m_{Pl}/\sqrt{\xi}$ where the Higgs field contributions to processes are absent \cite{Bezrukov:2008ej}. Here we adopt a somewhat pragmatic approach and  following in the spirit of reference \cite{DeSimone:2008ei}, we employ the suppressed Higgs effective standard model, now with the inclusion of the dark matter scalar, and explore the dependence of the measured cosmological density spectrum parameters on the Higgs and dark matter coupling constants for the inflationary scales $h  \sim m_{Pl}/\sqrt{\xi}$,  albeit possibly at the limits of its domain of validity.  Hence, Feynman integrands involving Higgs propagators will have a suppression factor $s$ for each such line with $s(h)$ evaluated at the scale of the background Higgs field $h$.  In particular the renormalization group functions will include a suppression factor for each physical Higgs line contributing to the function.

Isolating the explicit dependence on the dominant large Higgs field background terms, while the ellipses refer to the remaining subdominant singlet and standard model terms, the one-loop renormalization group improved effective action takes the form
\bwt
\be
\Gamma = \int d^4x \sqrt{-g}\left[\Lambda +\frac{1}{2}m_{Pl}^2 {f}(t) R +\frac{1}{2}{G^2}(t) \partial_\mu h g^{\mu\nu}\partial_\nu h -{V}(t)+\cdots\right].
\ee
\ewt
In the inflationary region, the Higgs-inflaton field has a large background value and the dominant forms of the renormalization group improved effective action coefficients are 
\bea
{V} (t) &=& \frac{{\lambda} (t)}{4}G^4(t) h^4 (t) = \frac{{\lambda} (t)}{4}G^4(t)m_t^4 e^{4t} \cr
{f} (t) &=& 1 + {\xi}(t) G^2 (t) \frac{h^2 (t)}{m_{Pl}^2} =1+ \xi(t) G^2(t) \frac{m_t^2}{m_{Pl}^2} e^{2t} , \cr
 & & 
\eea
where 
\be
G(t) = e^{-\int_0^t dt^\prime \gamma (t^\prime)},
\ee
with $\gamma (t)$ the Higgs field anomalous dimension.  Here we have introduced the scaling variable $t=t(h)=\ln{({h}/{m_t})}$ so that $h (t) =m_t e^t$ and have normalized the Higgs field and all running couplings at the top quark mass. The various $t$ dependent running coupling constants of the effective interaction terms are given by their renormalization group equations so that, for example,
\be
\frac{d{\lambda}(t)}{dt}=\beta_\lambda (t), \qquad  \qquad 
\frac{d{\xi}(t)}{dt}=\beta_\xi (t) .
\ee   
The one-loop $\beta$-functions and Higgs field anomalous dimension including the suppression factors are compiled in Appendix \ref{appendixA}.  

Cosmological quantities can most readily be calculated in the Einstein frame which is obtained by  rescaling the metric by $g_{E\mu\nu} = {f}(t) g_{\mu\nu}$.
The one-loop renormalization group improved effective action then takes the form 
\bea
\Gamma &=& \int d^4x \sqrt{-g_E}\left[\Lambda_E +\frac{1}{2}m_{Pl}^2 R_E  \right. \cr
 & &\left. +\frac{1}{2}\frac{1}{{f (t)} {\bar{s} (t)}} \partial_\mu h g_E^{\mu\nu}\partial_\nu h -{V}(t)/{f (t)}^2+\cdots\right],\cr
 & &
\eea
where
\be
\bar{s} (t) = \frac{f(t)}{G^2 (t) f(t) + \frac{3}{2} m_{Pl}^2 f^{\prime 2}(t)}.
\ee
Defining a canonically normalized field $\sigma = \sigma(h)$, the Einstein frame effective action reads
\bea
\Gamma &=& \int d^4x \sqrt{-g_E}\left[\Lambda_E +\frac{1}{2}m_{Pl}^2 R_E \right. \cr
 & &\left. \quad\quad +\frac{1}{2} \partial_\mu \sigma g_E^{\mu\nu}\partial_\nu \sigma -{V}_E (\sigma ) +\cdots\right],
\eea
where the dominant part of the one-loop renormalization group improved effective potential in the inflationary region in the Einstein frame is
\be
{V}_E = \frac{{V}(t)}{{f}^2(t)}
\ee
with the canonically normalized Higgs-inflaton field defined through
\be
\left(\frac{d\sigma}{dh}\right)^2 = \frac{1}{{f}\bar{s}}=\frac{{G^2} {f} + \frac{3}{2}m_{Pl}^2 {f}^{\prime 2}}{{f}^2} .
\ee

A Higgs field wavefunction renormalization suppression factor is now defined as
\bea
Z(t) &\equiv & G^2 (t) \bar{s} (t)\cr
&=&\frac{G^2(t) {f}}{G^2(t) {f} + \frac{3}{2}m_{Pl}^2 {f}^{\prime 2}}\cr
 &=&\frac{1+{\psi}^2 (t)}{ \left(1+[1+6\xi (t)(1-\gamma (t)+ \beta_\xi /2\xi (t))^2 ]{\psi}^2(t)\right)} ,\cr
 & & 
\eea
where 
\be
\psi^2 (t) = \xi (t) G^2 (t) h^2 (t) /m_{Pl}^2 =\xi (t) G^2 (t) e^{2t}m_t^2/m_{Pl}^2
\ee
is a  renormalization group invariant dimensionless field. Note that ${f}(t) = 1+ \psi^2 (t)$. As  a function of $\psi(t)$, the Higgs-inflaton effective potential takes the simple form
\be
{V}_E = \frac{m_{Pl}^4}{4}\frac{\lambda (t)}{\xi^2 (t)}\frac{\psi^4 (t)}{(1+\psi^2 (t))^2} .
\ee

The renormalization group invariant slow roll inflationary parameters are defined by (once again prime denotes differentiation with respect to $h$) 
\bea
\epsilon &=&\frac{1}{2}m_{Pl}^2 \left(\frac{1}{{V}_E}\frac{d{V}_E}{d\sigma}\right)^2 =\frac{1}{2}m_{Pl}^2 \left(\frac{{V}_E^\prime}{{V}_E}\right)^2 \left(\frac{d\sigma}{dh}\right)^{-2} \cr
 &=&8\xi \frac{Z}{{f} \psi^2}\left(1-\gamma +\frac{\beta_\lambda}{4\lambda}  +\psi^2 \left( \frac{\beta_\lambda}{4\lambda} - \frac{\beta_\xi}{2\xi}\right)\right)^2\cr
\eta &=& m_{Pl}^2 \frac{1}{{V}_E}\frac{d^2{V}_E}{d\sigma^2}
= 2\epsilon +m_{Pl}\frac{1}{\sqrt{2\epsilon}}\frac{d\epsilon}{dh}\left(\frac{d\sigma}{dh}\right)^{-1}\cr
 &=& 2\epsilon +\frac{\sqrt{\xi {f} Z}}{\sqrt{2 \epsilon}}\left(\frac{1}{\psi} \right)\frac{d\epsilon}{dt}\cr
\zeta^2 &=& m_{Pl}^4 \frac{1}{{V}_E}\frac{d^3{V}_E}{d\sigma^3}\frac{1}{{V}_E}\frac{d{V}_E}{d\sigma}
= 2\eta \epsilon + m_{Pl} \sqrt{2\epsilon} \frac{d\eta}{dh}\left(\frac{d\sigma}{dh}\right)^{-1}\cr
 &=& 2\eta \epsilon + \sqrt{2 \epsilon \xi {f} Z} \left(\frac{1}{\psi} \right)\frac{d\eta}{dt},
\label{slowrollparameters}
\eea
where all quantities are evaluated at the onset of inflation.
The power spectrum of density perturbations in $k$-space is given by 
\bea
P_s(k)= \Delta^2_{{\cal R}} \biggr(\frac{k}{k^*}\biggr)^{n_s(k)-1},
\eea
where the amplitude of density perturbations is expressed as
\be
\Delta_{{\cal R}}^2 = \frac{{V_E}}{24 \pi^2 m_{Pl}^4 \epsilon}\biggr\vert_{k^*}
\label{amp}
\ee
and is secured by the combination of experimental results from WMAP5, Baryon Acoustic Oscillations (BAO) and supernovae (SN) as: $\Delta^2_{{\cal R}} =(2.445 \pm 0.096)\times 10^{-9}$ at $k^*=0.002$ Mpc$^{-1}$ \cite{Hinshaw:2008kr}.  Slow roll inflation predicts the spectral index, $n_s$, its running, $\alpha =dn_s/d\ln{k}$, and the tensor to scalar ratio, $r$, to be
\bea
n_s &=& 1-6\epsilon + 2 \eta \cr
\alpha &=& -24 \epsilon^2 +16 \epsilon \eta -2 \zeta^2 \cr
r &=& 16 \epsilon .
\label{spectral}
\eea
The WMAP5, BAO and SN experimental evidence gives $n_s=0.960 \pm 0.013$ and $r<0.22$ (95\% CL) with an insignificant running spectral index, $\alpha =-0.028\pm 0.020$ \cite{Hinshaw:2008kr}.

\section{Numerical Results \label{section3}}

The renormalization group equations for the various running couplings are solved numerically starting from  $t=0$ which corresponds to the top quark mass. The top Yukawa coupling at $t=0$ is fixed by the central value deduced from the top quark mass $m_t = {y}_t (0) v/\sqrt{2} \equiv 171.2$ GeV , while the gauge coupling constants are normalized at $m_t$ as $\alpha_1 (0)=0.0102718$, $\alpha_2 (0)=0.0334412$ and $\alpha_3 (0)=0.108635$ \cite{Amsler:2008zzb}. The value of $\xi(0)$ is determined so that at the initial point of the slow roll inflation, $t_i$, the non-minimal coupling constant $\xi(t_i)$ is such that the calculated value of the amplitude of density perturbations, equation (\ref{amp}), agrees with the measured result.  The Higgs-inflaton exits inflation at the final point $t_f$ when the first slow roll parameter is 1: $\epsilon (t_f) =1$.  The number of e-folds of expansion between $t_i$ and $t_f$ is 
\bea
N_e (t_i) &=& \frac{1}{\sqrt{2}m_{Pl}}\int_{\sigma_{f}}^{\sigma_i} \frac{d\sigma^\prime}{\sqrt{\epsilon(h (\sigma^\prime))}}\cr
 &=&\frac{1}{\sqrt{2}m_{Pl}}\int_{h_{f}}^{h_i} \frac{dh^\prime}{\sqrt{\epsilon(h^\prime)}}\left(\frac{d\sigma^\prime}{dh^\prime}\right)\cr
&=&\int_{t_{f}}^{t_{i}} dt \frac{\psi^2}{4{\xi} Z}\frac{1}{\biggr(1-\gamma+\frac{\beta_\lambda}{4\lambda}+\psi^2(\frac{\beta_\lambda}{4\lambda}-\frac{\beta_{\xi}}{2{\xi}})\biggr)}.\cr
 & &  
\eea
The initial point of inflation is defined so that $N_e (t_i)=60$.  These three criteria are then implemented recursively for each choice of initial Higgs mass, $m_h =\sqrt{2 \lambda (0)}~v$ and scalar coupling constants $\kappa (0)$, $\lambda_S (0)$ and $\xi_S (0)$.  The standard model without dark matter coupling corresponds to $\kappa (0) =0=\lambda_S (0)$.

\begin{figure}
\begin{center}
\includegraphics[scale=0.95]{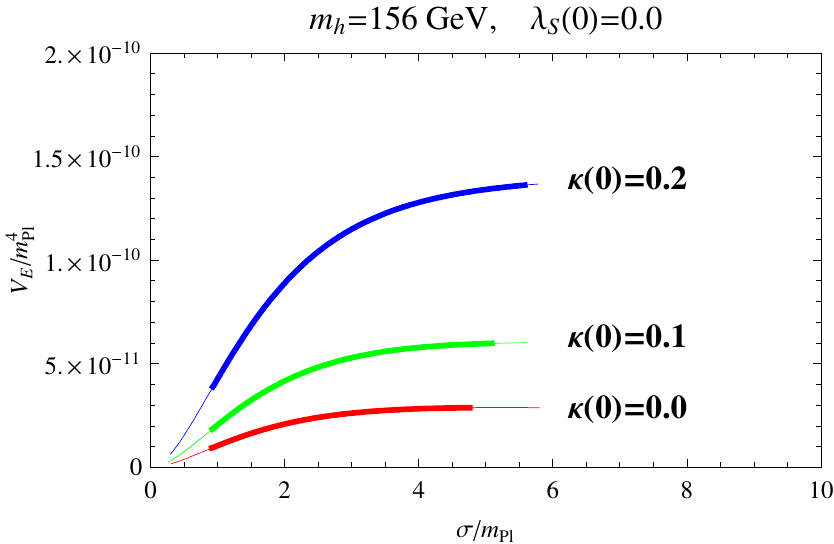}
\end{center}
\caption{The Einstein frame renormalization group improved effective potential as a function of the canonically normalized Higgs-inflaton field.  The magnitude and shape of this potential in the inflationary cosmological state varies with the strength of the Higgs-inflaton and dark matter coupling constant $\kappa$.  The thickened portion of the potential curve corresponds to the $N_e =60$ e-folds of inflation with onset and exit values of $\sigma$ as shown.}
\label{EffPotential-1}
\end{figure}
\begin{figure}
\begin{center}
\includegraphics[scale=0.75]{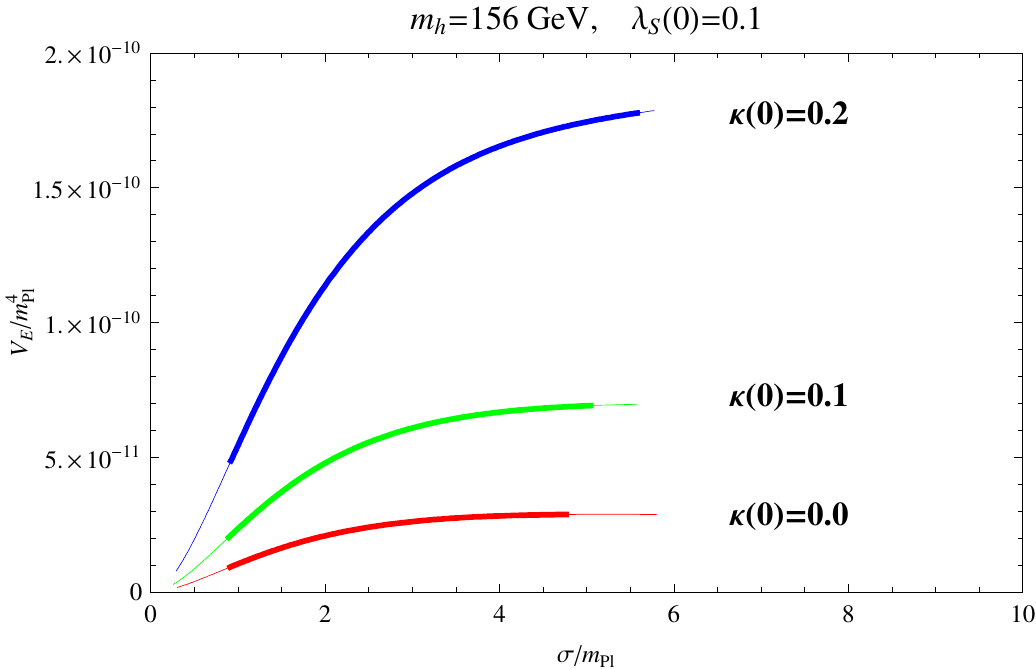}
\end{center}
\caption{The Einstein frame renormalization group improved effective potential as a function of the canonically normalized Higgs-inflaton field.  The magnitude and shape of this potential in the inflationary cosmological state varies with the strength of the dark matter self-coupling constant $\lambda_S$ as compared to Fig. 1 for the different Higgs-inflaton to dark matter coupling $\kappa$.  The thickened portion of the potential curve corresponds to the $N_e =60$ e-folds of inflation with onset and exit values of $\sigma$ as shown.}
\label{EffPotential-2}
\end{figure}

\begin{figure}
\begin{center}
\includegraphics[scale=0.71]{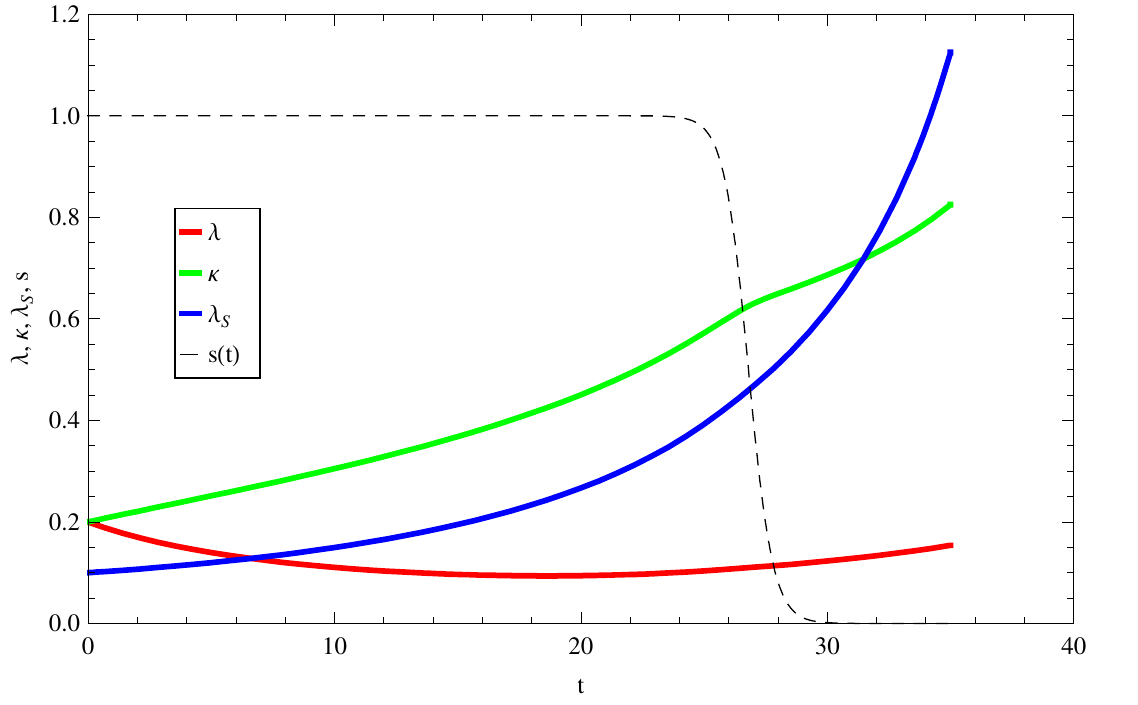}
\end{center}
\caption{The running coupling constants for the scalar fields.  The initial conditions for the coupling constants correspond to the effective potential plot of Fig. \ref{EffPotential-2} with $\kappa (0) =0.2$, $\xi (0)=8,315$ and $\xi_S (0) =0.0$.  In this case the onset of inflation occured at the scale $t_i=35$ with exit at $t_f=32.7$ after 60 e-folds of expansion.}
\label{EffCoupling-Suppressed-1}
\end{figure}

\begin{figure}
\begin{center}
\includegraphics[scale=0.75]{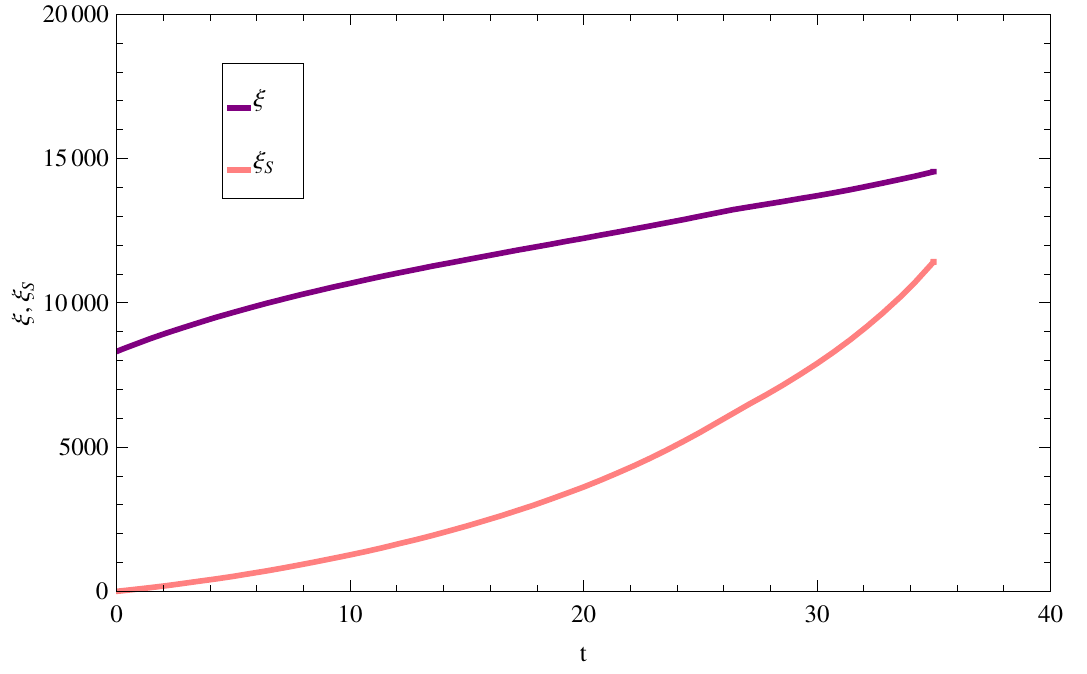}
\end{center}
\caption{The running of the non-minimal gravitational coupling constants for the scalar fields.  The initial conditions for the coupling constants correspond to the effective potential plot of Fig. \ref{EffPotential-2} with $\kappa (0) =0.2$, $\xi (0)=8,315$ and $\xi_S (0) =0.0$.  In this case the onset of inflation occured at the scale $t_i=35$ with exit at $t_f=32.7$ after 60 e-folds of expansion.}
\label{EffCoupling-Suppressed-2}
\end{figure}

As evident from Figs. \ref{EffPotential-1} and \ref{EffPotential-2}, the shape of the effective potential in the inflation region depends not only on the Higgs boson mass \cite{DeSimone:2008ei,Bezrukov:2008ej,Barvinsky:2009fy}, but also changes as the Higgs-inflaton to dark matter coupling and dark matter self-coupling varies.  The high energy suppression of the Higgs propagator also effects the running of the scalar coupling constants as seen in Figs. \ref{EffCoupling-Suppressed-1} and \ref{EffCoupling-Suppressed-2}.  The suppression factor is plotted along with the coupling constants whose $\beta$-function dependence on the suppression factor manifests itself in the abrupt change of running.

\begin{figure*}
$\begin{array}{cc}
  \includegraphics[scale=1.0]{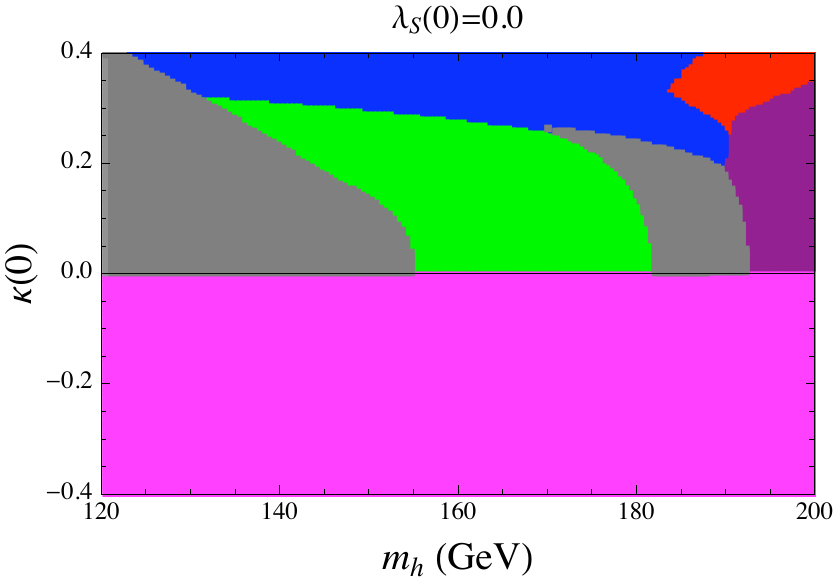}&
  \includegraphics[scale=1.0]{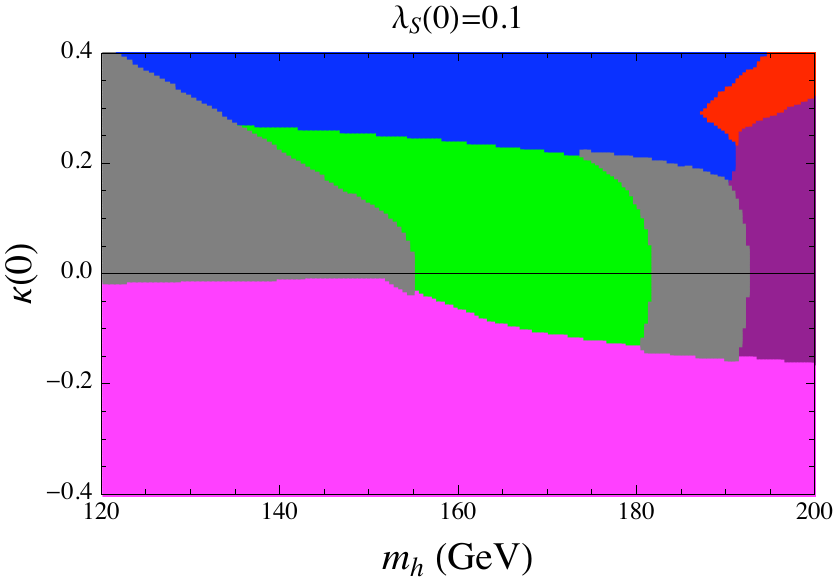}\\
  \includegraphics[scale=1.0]{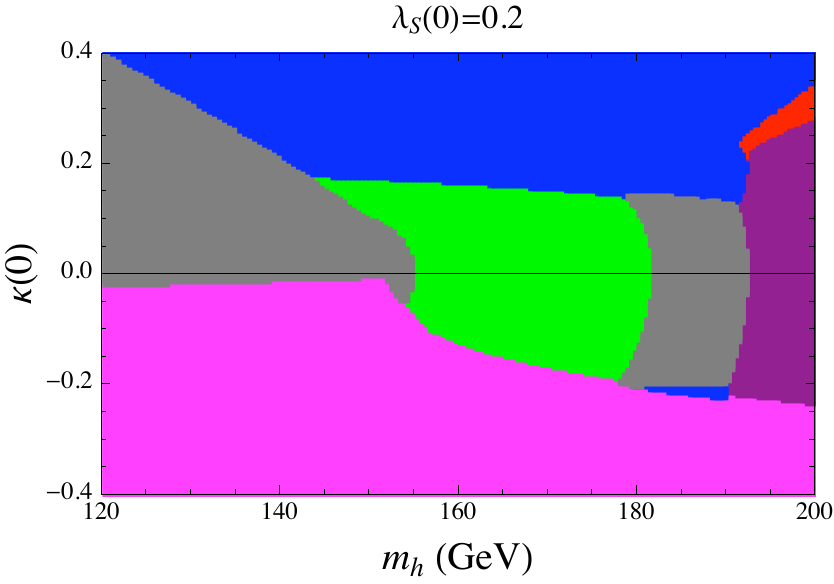}&
  \includegraphics[scale=1.0]{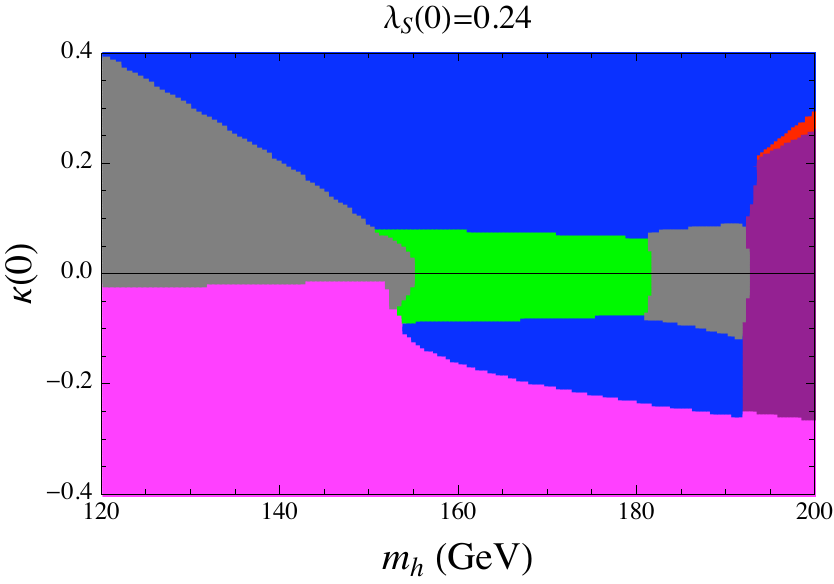}
\end{array}$
\begin{center}
\includegraphics[scale=1.0]{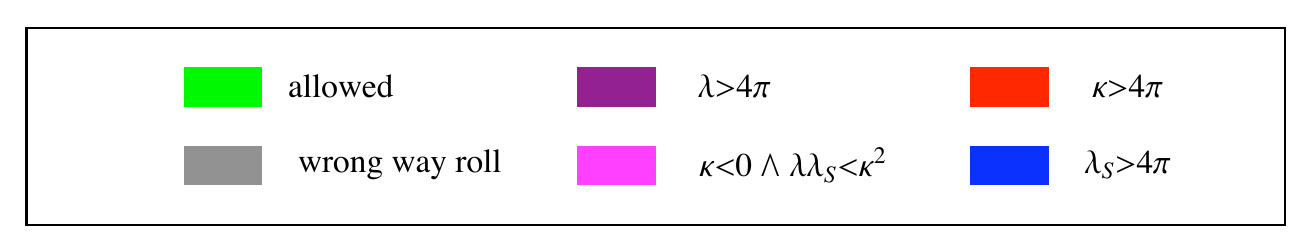} 
\end{center}
\caption{The wrong way roll constraints for parameter space are added to those of vacuum stability and triviality (compare to Fig. \ref{suppression}).  These are displayed for typical initial non-minimal gravitational couplings of $\xi (0) = 10^4$ and $\xi_S (0) =0.0$.  The grey colored areas mark the wrong way roll excluded regions of parameter space.  The constraints apply to scales up to those typical of the onset of inflation, $t_i=34.5$.}
\label{NoRollConstraints}
\end{figure*}

The fractional rate of change of the effective potential is obtained from the renormalization group equations.  Recalling that
\bea
\frac{d}{dt}\ln{\frac{\lambda}{\xi^2}} &=& 4\left( \frac{\beta_\lambda}{4\lambda}-\frac{\beta_\xi}{2\xi}\right) \cr
\frac{d}{dt}\ln{\frac{\psi^4}{(1+\psi^2)^2}}&=& 4\left( \frac{1-\gamma +\frac{\beta_\xi}{2\xi}}{(1+\psi^2)}\right) ,
\eea
it follows that 
\be
\frac{d}{dt}\ln{{V}_E}= \frac{4}{(1+\psi^2)}\left(1-\gamma +\frac{\beta_\lambda}{4\lambda}+\psi^2  \left[ \frac{\beta_\lambda}{4\lambda}-\frac{\beta_\xi}{2\xi}\right] \right).
\ee
For larger values of $\lambda (0)$ and hence more massive Higgs bosons, the term in the square bracket can cancel the remaining bracketed terms since $\psi^2$ is getting large. Thus there will be some $t$ value at which the effective potential  reaches a maximum value, the wrong way roll point, before turning over and decreasing as the scale is increased.  For a Higgs-inflaton model to be viable,  it is thus necessary that the inflation must initiate  at $t_{i}$ values less than that of this relative maximum  or else the inflaton rolls to $m_{Pl}$ and beyond. The scale at which this cannot be accomplished sets an upper limit on the Higgs mass consistent with the inflationary cosmology model. It is typically below that of the  suppressed triviality bound which is mapped out in Appendix \ref{appendixB}.  Hence, an additional region of parameter space previously allowed by the suppressed triviality constraint, as seen in Fig. \ref{suppression}, will be ruled out due to the wrong way roll condition.  Likewise, for smaller values of $\lambda (0)$ that lead to a running close to the vacuum stability bound, $\beta_\lambda / \lambda$ is negatively large enough so that a relative maximum forms in the effective potential.  Thus once again, there is a region of parameter space not ruled out by vacuum stability but excluded by the wrong way roll condition.  Moreover, since the effective potential has a maximum before vacuum instability and the onset of inflation are reached, the wrong way roll excluded area covers that previously ascribed to vacuum instability.  The wrong way roll criterion more severely restricts the parameter space of the model than the requirement of absolute vacuum stability up to the onset of inflation.  For that matter, even if the absolute stability constraint is abandoned in favor of vacuum metastability \cite{Espinosa:2007qp}  with a lifetime longer than the age of the observable universe, then the additional range of Higgs boson masses allowed by this less stringent condition are ruled out by the imposition of the wrong way roll constraint. Hence for both low and high Higgs masses, additional regions of parameter space are excluded as they do not support slow roll inflation.  For typical values of the scalar parameters the wrong way roll excluded region of parameter space is displayed in Fig. \ref{NoRollConstraints}.

For $\kappa(0)=\lambda_S(0)=0$, the allowed range of Higgs boson masses after the imposition of the wrong way roll constraint is roughly 155 GeV $< m_h < $ 182 GeV. This is a slightly smaller range than that allowed without the additional constraint which is 153 GeV $< m_h < $ 190 GeV (see Appendix B). Still considering $\lambda_S(0)=0$, as $\kappa(0)$ increases, the smallest allowed $m_h$ value consistent with the various constraints is approximately 130 GeV, which is roughly the same as without the wrong way roll constraint. This occurs when $\kappa(0)\simeq 0.3$. For larger values of $\kappa(0)$, the allowed parameter space vanishes. Once again, this is akin to the case without the wrong way roll constraint. As $\lambda_S(0)$ increases from zero, the allowed parameter space starts to shrink as a smaller range of $\kappa(0)$ values are permitted, while there remains a finite range of allowed Higgs boson masses.  Finally, for $\lambda_S(0) >0.25$, the allowed $\kappa(0)$ range vanishes for a finite range of  allowed Higgs boson masses. Hence the allowed parameter space disappears. While Fig. \ref{NoRollConstraints} corresponds to particular value of $\xi(0)$, the generic features continue to hold for a range of large $\xi(0)$ values. 

\begin{figure*}
$\begin{array}{cc}
\includegraphics[scale=0.95]{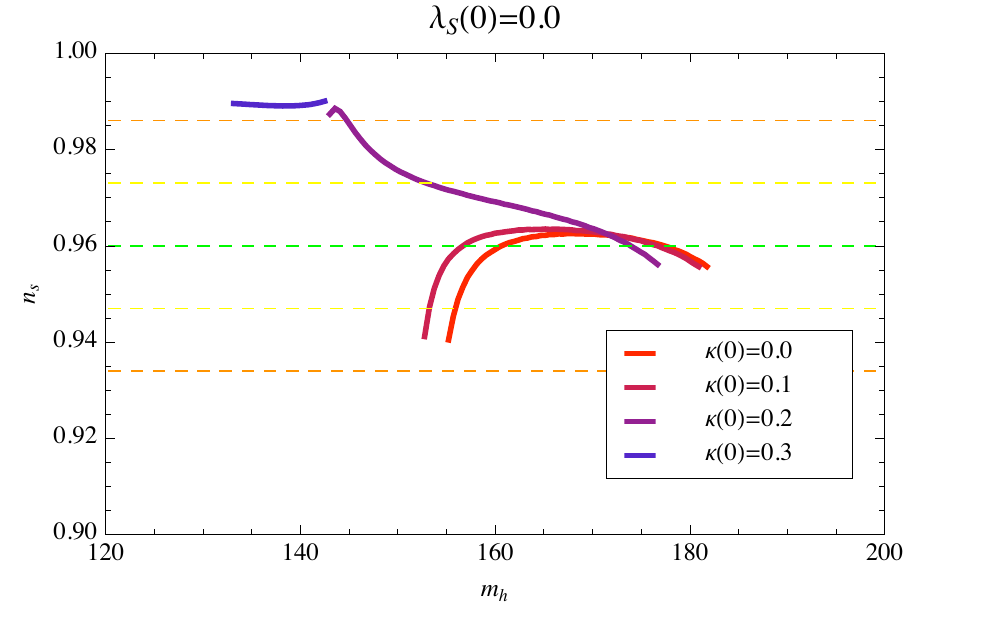} &
\includegraphics[scale=0.90]{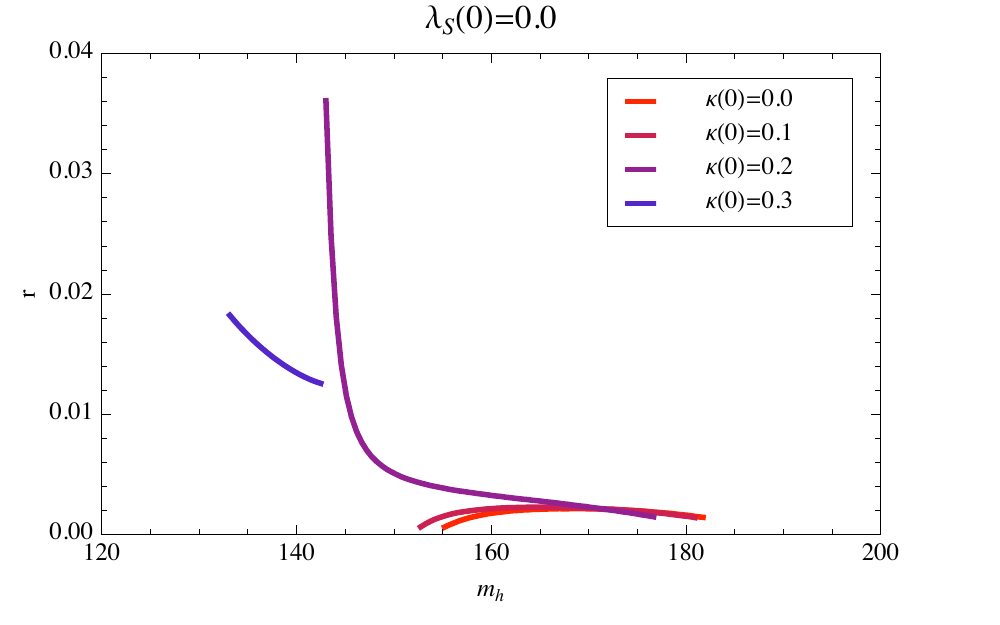} \\
\includegraphics[scale=0.90]{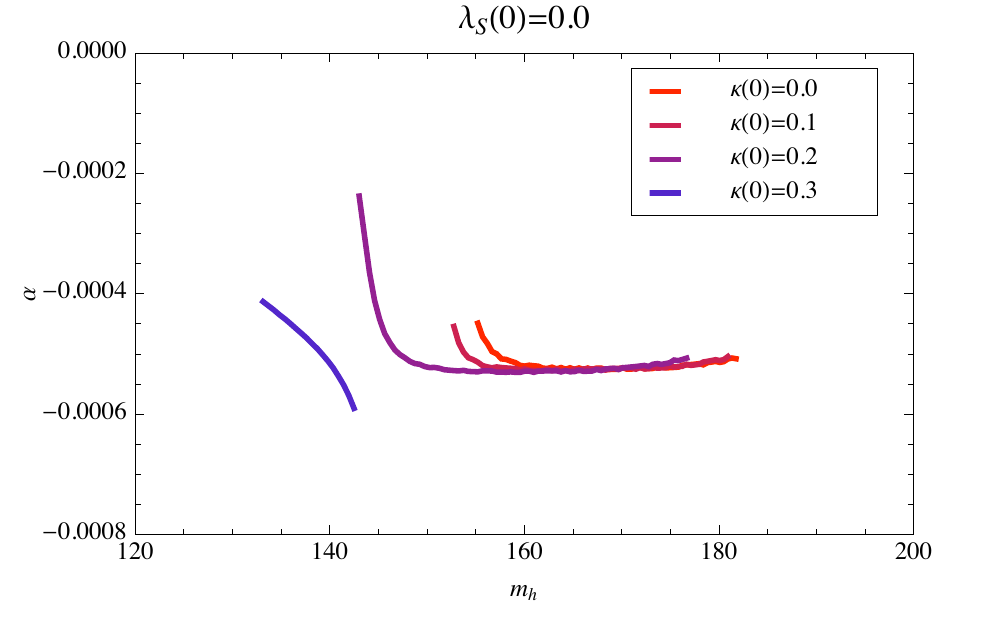}
\end{array}$
\caption{The spectral index $n_s$, its running $\alpha$ and the tensor to scalar ratio $r$ are plotted against the Higgs mass $m_h$ for various values of the Higgs-inflaton to dark matter coupling constant $\kappa (0)$ for the fixed initial value of the dark matter self-coupling $\lambda_S (0) =0.0$.  Curve endpoints are determined by the wrong way roll, triviality and vacuum stability conditions. The dashed horizontal lines in the spectral index plot denote its central value, 0.960, and one and two standard deviations from it.}
\label{CosmoQ-1}
\end{figure*}

\begin{figure*}
$\begin{array}{cc}
\includegraphics[scale=0.95]{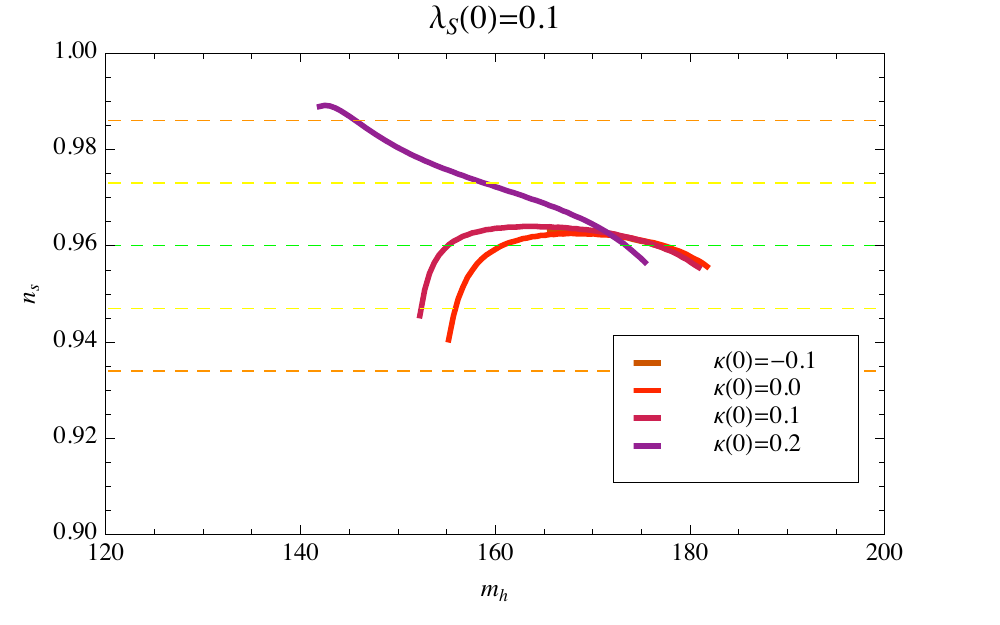} &
\includegraphics[scale=0.90]{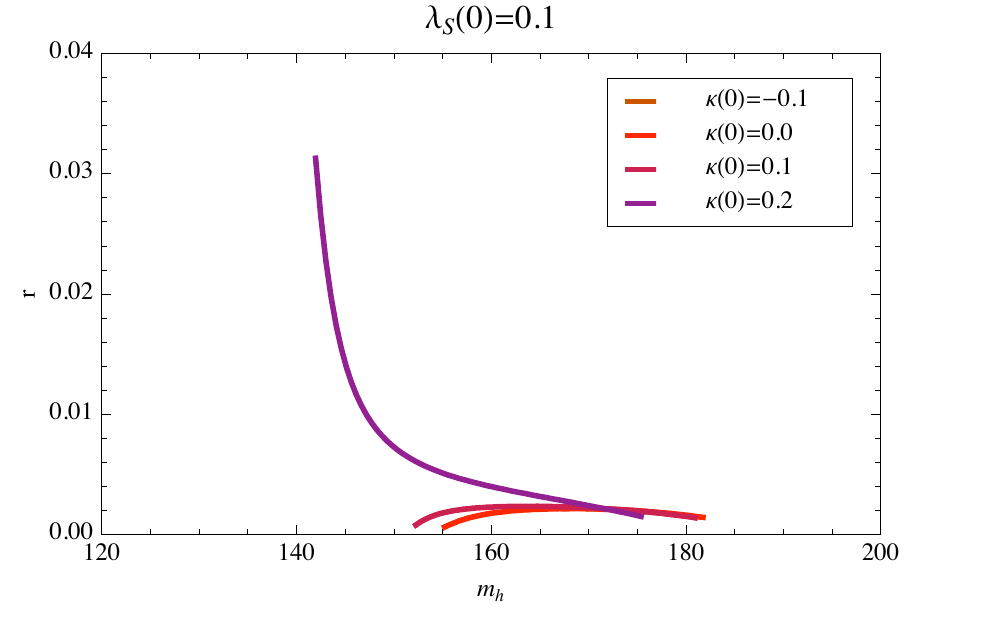} \\
\includegraphics[scale=0.90]{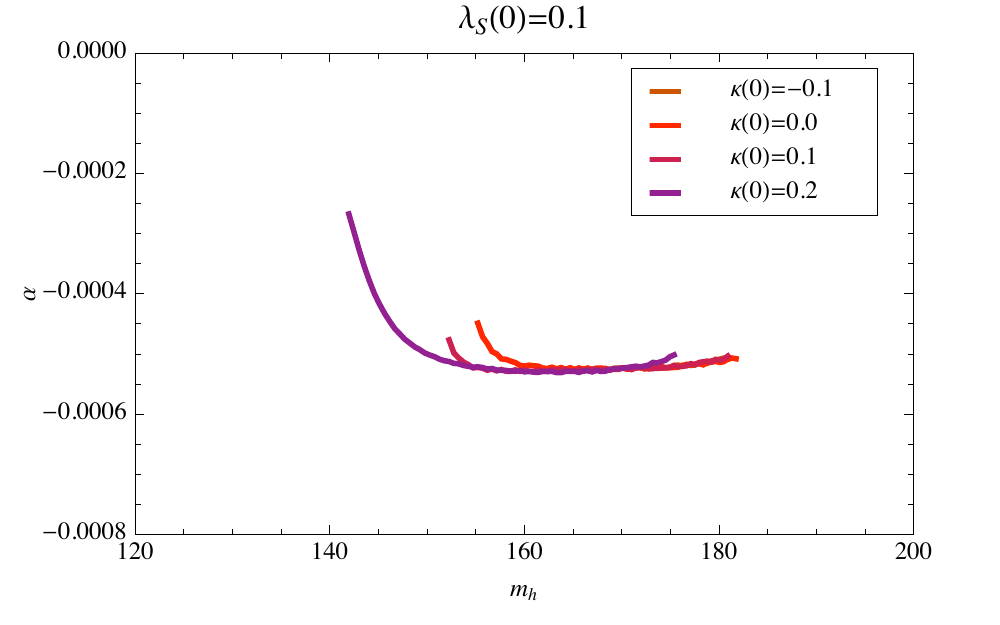}&
\end{array}$
\caption{The spectral index $n_s$, its running $\alpha$ and the tensor to scalar ratio $r$ are plotted against the Higgs mass $m_h$ for various values of the Higgs-inflaton to dark matter coupling constant $\kappa (0)$ for the fixed initial value of the dark matter self-coupling $\lambda_S (0) =0.1$.  Curve endpoints are determined by the wrong way roll, triviality and vacuum stability conditions.  The dashed horizontal lines in the spectral index plot denote its central value, 0.960, and one and two standard deviations from it.}
\label{CosmoQ-2}
\end{figure*}

\begin{figure*}
$\begin{array}{cc}
\includegraphics[scale=1.0]{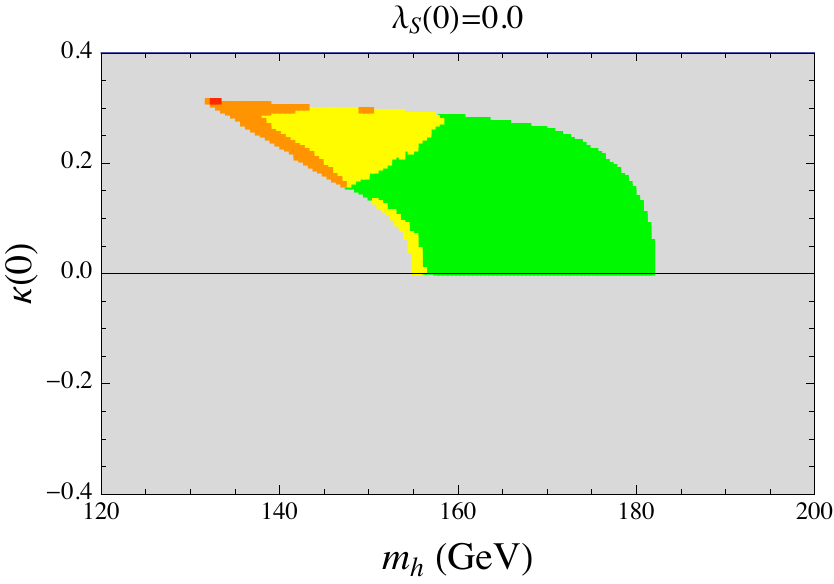}&
  \includegraphics[scale=1.0]{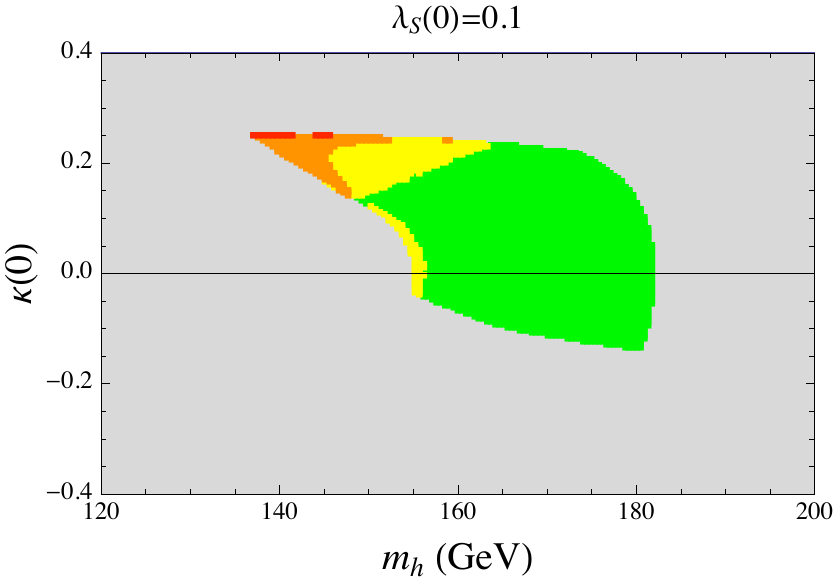}\\
  \includegraphics[scale=1.0]{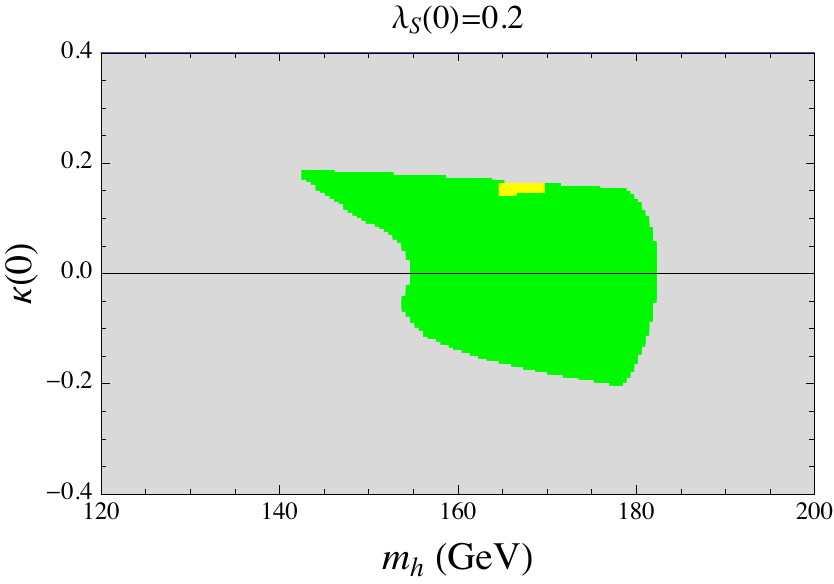}&
  \includegraphics[scale=1.0]{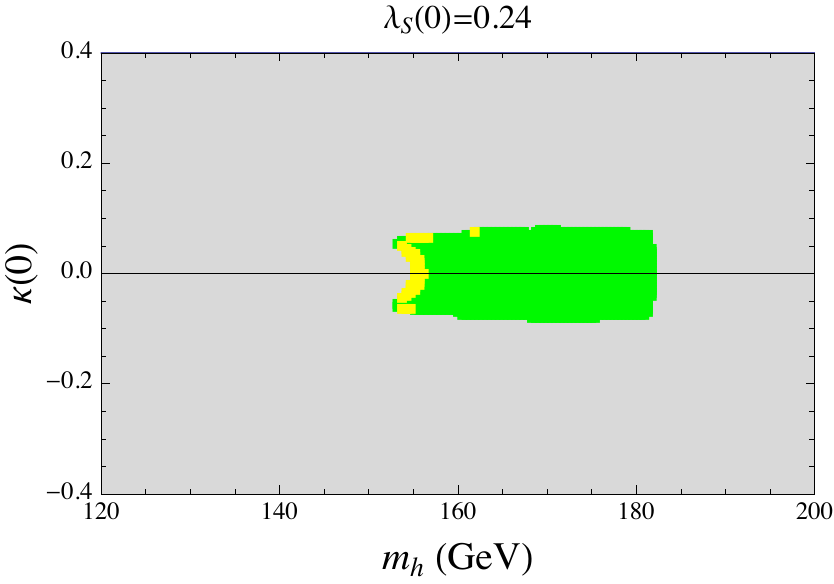}
\end{array}$
\caption{The spectral index cosmological constraints on slices of parameter space for different dark matter self coupling $\lambda_S$ as determined by the degree of agreement with the experimental value of the spectral index, $n_s=0.960$.  These constraints are applied to the potentially allowed (green) areas displayed in Fig. \ref{NoRollConstraints}.  Here the green regions indicate the volume of parameter space that predicts spectral index values within one standard deviation of the central value.  The yellow regions correspond to calculated values between one and two standard deviations of the central value, while the orange regions correspond to two to three standard deviations from it.  The red areas indicate parameters that predict spectral index values more than three standard deviations from the central value.  Finally the grey region is excluded by triviality and vacuum stability bounds along with the wrong way roll condition.  For $\lambda_S (0) \ge 0.25$, there is no allowed region of parameter space.}
\label{CosmoParameterSpace-00}
\end{figure*}

The viable region of parameter space is further restricted by the requirement of yielding a value of the spectral index within its experimentally determined range.  Once the effective potential is determined, the cosmological quanities are calculated using equations (\ref{slowrollparameters}) and (\ref{spectral}).  The spectral index, its running and the tensor to scalar ratio are plotted versus the Higgs boson mass in Figs. \ref{CosmoQ-1} and \ref{CosmoQ-2} for various initial values of the Higgs-inflaton to dark matter coupling and for two initial values of the dark matter self-coupling.  The green area shown in Fig. \ref{CosmoParameterSpace-00} corresponds to model values of $n_s$ that are within one standard deviation of the experimental central value for the spectral index, $n_s=0.960$ \cite{Hinshaw:2008kr}.  The yellow area is determined by the predicted value of $n_s$ being between one and two standard deviations of its central value.  The orange regions correspond to $n_s$ predicted values to be within two to three standard deviations from the central value.  Finally the red regions indicate a predicted spectral index greater than three standard deviations beyond the central value.  The grey regions are excluded by triviality, vacuum stability and the wrong way roll conditions.  

In general, as the Higgs boson to dark matter coupling strength, $\kappa (0)$, increases, lower values of the Higgs mass will still support a stable vacuum as well as avoid the wrong way roll condition.  However, a tension begins to arise between the experimentally allowed values of the spectral index and the model values of the index at low Higgs mass.  It follows from  Figs. \ref{CosmoQ-1}, \ref{CosmoQ-2}, and \ref{CosmoParameterSpace-00} that, consistent with the triviality, vacuum stability and wrong way roll constraints, agreement with the central measured value of $n_s$ favors a Higgs boson mass in the range 155-180 GeV and a smaller value of $\kappa(0)$. As $\kappa(0)$  grows to values greater than $\simeq 0.3$,  the computed value of $n_s$ lies between one to three standard deviations above the central measured value and occurs for smaller Higgs boson masses of order 130-145 GeV. Thus a discovery of a Higgs boson mass in this range favors a larger coupling to dark matter as is also preferred by the dark matter abundance calculations provided the dark matter is either lighter or heavier than $\sim m_h/2$. Note that there is no additional constraint arising from  $r$ and $\alpha$ as their computed values lie well below the present experimental limits.

As noted previously, the parameter space available shrinks as the  dark matter self coupling increases as displayed in Fig. \ref {CosmoParameterSpace-00} for different $\lambda_S(0)$ slices.  This self coupling is largely unconstrained by dark matter experiments and observations. For small dark matter mass (less than $\sim 1$ GeV) a correlation between the self coupling and mass may soften current discrepancies between the observed dark matter halo structure and numerical simulations of the structure formation process \cite{Bento:2000ah,McDonald:2001vt}.  Overall there is ample parameter space available for a consistent minimal standard model with the Higgs boson acting as the inflaton and interacting with scalar dark matter. 
\clearpage

\section{Conclusions \label{section4}}

The influence of the inclusion of dark matter on slow roll inflation models where the inflaton is identified with the standard model Higgs boson was explored.  To achieve this, the standard model was modified by the inclusion of an  hermitian scalar standard model singlet field, which accounts for the observed abundance of dark matter and the lack of its direct detection, and a large non-minimal coupling of the Higgs doublet to the Ricci scalar curvature. The model parameter space thus included the Higgs boson and singlet scalar masses, the scalar self coupling and the  coupling between the Higgs doublet and the dark matter in addition to the non-minimal gravitational couplings of both the Higgs doublet and the singlet scalar. In the inflationary region where the physical Higgs field develops a sizeable classical background, the presence of the large Higgs doublet non-minimal gravitational coupling results in a highly suppressed  physical Higgs field propagator. Accounting for this, the one-loop renormalization group improved effective potential was computed and the constraints on the model parameter space were delineated. In addition to the usual triviality and vacuum stability bounds, we focused on the cosmological constraints arising from the identification of the Higgs boson with the inflaton. Since, in general, the one loop effective potential develops a maximum, it was necessary to insure that the onset of inflation occurred such that the inflaton rolling was toward the origin and not towards the Planck scale. The implication of this wrong way roll constraint is displayed in Fig. \ref{NoRollConstraints} and is seen to eliminate even more of the parameter space than the vacuum stability (or metastability) constraint. Various parameters characterizing the slow roll inflation were computed with the spectral index, $n_s$, providing the most stringent constraint on the coupling constant space. As is seen in Fig. \ref{CosmoParameterSpace-00}, the region of parameter space allowed after the imposition of wrong way roll, triviality and vacuum stability constraints, is further partitioned into various sections whose agreement with the measured spectral index is only at a varying number (one - three) of standard deviations above the central value.  Larger values of the coupling of the Higgs-inflaton to dark matter, as is preferred by the dark matter abundance calculation, lead to a lower allowed range of Higgs boson masses.  In addition, there are the various Higgs mass bounds from accelerator experiments.  The direct LEP search \cite{Barate:2003sz} gives a lower bound of 114.4 GeV, while a fit to electroweak precision data prefers \cite{Grunewald:2007pm} a mass $m_h <$ 182 GeV.  Finally, a combined CDF and DO analysis of Tevatron data \cite{Benjamin:2009bn} excludes the range 160 GeV $< m_h <$ 170 GeV.  While our analysis has been limited to the one-loop radiative corrections, we anticipate that the general features of our results will continue to hold when two and higher loop corrections are included.  Thus we conclude that even after the inclusion of the dark matter, there still remains a range of LHC attainable Higgs boson mass values that are consistent with the cosmological parameters of slow roll inflation when the Higgs scalar is identified as the inflaton. 

\begin{acknowledgments}
The work of TEC, BL and STL was supported in part by the U.S. Department of Energy under grant DE-FG02-91ER40681 (Theory).  The work of TtV was supported in part by a Cottrell Award from the Research Corporation and by the NSF under grant PHY-0758073.  
\end{acknowledgments}
~\\

\appendix
\section{Renormalization Group Effective Coupling Constants \label{appendixA}} 

The one-loop renormalization group \cite{Barvinsky:1993zg} $\beta$-functions, including the Higgs field suppression factor $s(t)$, are 
\bwt
\bea
(4\pi)^2 \frac{d{g}_1}{dt} &=&{g}_{1}^3 \left(\frac{81+s(t)}{12}\right) \cr
(4\pi)^2\frac{d{g}_2}{dt} &=&-{g}_2^{3} \left(\frac{39-s(t)}{12}\right) \cr
(4\pi)^2\frac{d{g}_3}{dt} &=&-7{g}_3^3  \cr
(4\pi)^2\frac{d{y}_t}{dt} &=&{y}_t \left(\left(\frac{23}{6}+\frac{2}{3}s(t)\right){y}_t^2  -8{g}_3^2   -\frac{17}{12}{g}_1^2 -\frac{9}{4}{g}_2^2 \right) \cr
(4\pi)^2\frac{d\lambda}{dt} &=&\left((6+18s^2 (t)) \lambda^2 -6{y}_t^4+\frac{3}{8}\left(2{g}_2^4+({g}_1^2+{g}_2^2)^2\right) +12{y}_t^2 \lambda- 3{g}_1^2\lambda -9{g}_2^2 \lambda  +2\kappa^2 \right) \cr
(4\pi)^2\frac{d\kappa}{dt} &=&\kappa \left( 8 s(t)\kappa +6(1+s^2 (t)) {\lambda} +6{\lambda}_S +6{y}_t^2-\frac{3}{2}{g}_1^2-\frac{9}{2}{g}_2^2 \right) \cr
(4\pi)^2\frac{d{\lambda}_S}{dt} &=& 18~{\lambda}_S^2 +(6+2s^2 (t))~{\kappa}^2 \cr
(4\pi)^2\frac{d{\xi}}{dt} &=& \left({\xi}+\frac{1}{6}\right)\left( 6(1+s^2 (t))\lambda +6{y}_t^2 -\frac{3}{2}{g}_1^2-\frac{9}{2}{g}_2^2 \right)+\left({\xi}_S +\frac{1}{6}\right)~2\kappa\cr
(4\pi)^2\frac{d{\xi}_S}{dt}&=&\left({\xi}_S + \frac{1}{6}\right)~6{\lambda}_S + \left(\xi +\frac{1}{6}\right)(6+2s^2 (t))~\kappa .
\eea
\ewt
The Higgs field propagator suppression factor in the inflationary background, where $h (t) =m_t e^t$, is given by
\be
s(t) =\frac{1+\xi(t) m_t^2 e^{2t}/m_{Pl}^2}{1+(1+6\xi (t))~\xi (t) m_t^2 e^{2t}/m_{Pl}^2}
\label{s(t)}
\ee
and the Higgs field one loop anomalous dimension $\gamma$ is (in Landau gauge)
\be
(4\pi)^2 \gamma= 3{y}_t^2 -\frac{3}{4}{g}_1^2-\frac{9}{4}{g}_2^2 .
\ee

\section{Triviality and Vacuum Stability \label{appendixB}}

The effective coupling constants can be determined by integrating the one-loop renormalization group equations detailed in Appendix A.  The constraints due to absolute vacuum stability of the scalar sector provide lower bounds on the initial values of the coupling constants.  Vacuum stability requires that $\lambda \ge 0$, $\lambda_S \ge 0$ and for $\kappa < 0$ the relation $\lambda~\lambda_S \ge \kappa^2$ must hold for all values of $t$ in the range of applicability of the effective theory.  For the theory to remain in the perturbative sector of the model, the scalar coupling constants must not reach their respective Landau singularities for all values of $t$ in the range of applicability of the effective theory.  Thus we require the so-called triviality bounds that the coupling constants to be less than $4\pi$, $\lambda <4\pi$, $\lambda_S <4\pi$ and $\kappa <4\pi$. 

\begin{figure*}[ht]
$\begin{array}{cc}
\includegraphics[scale=1.0]{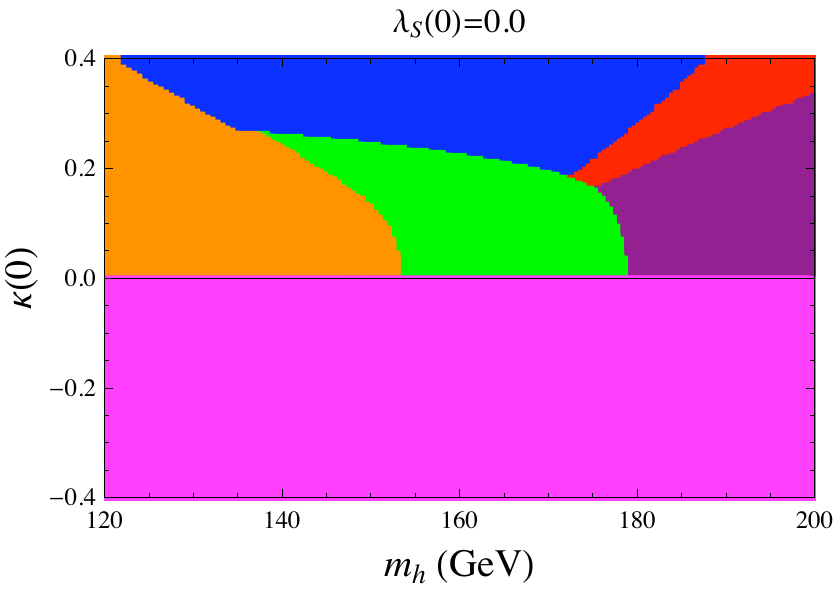} &
\includegraphics[scale=1.0]{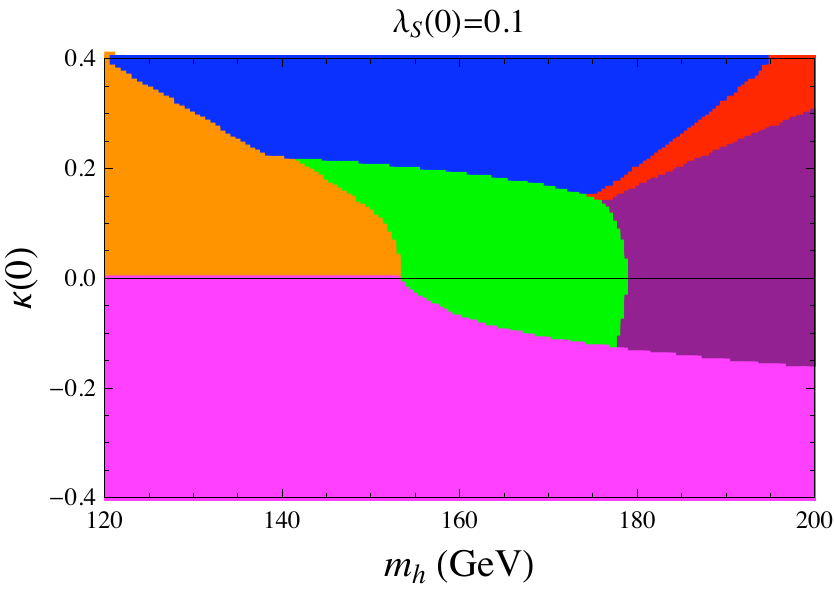} \\
\includegraphics[scale=1.0]{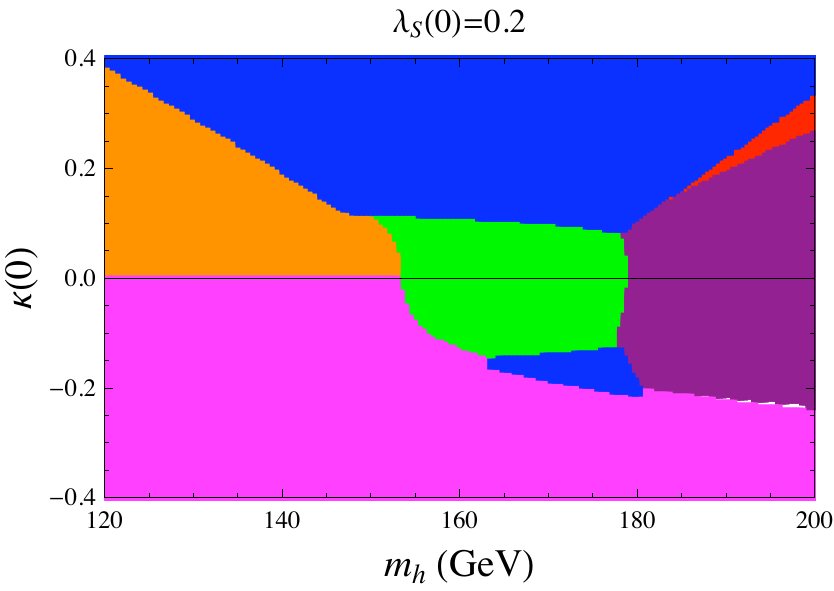} &
\includegraphics[scale=1.0]{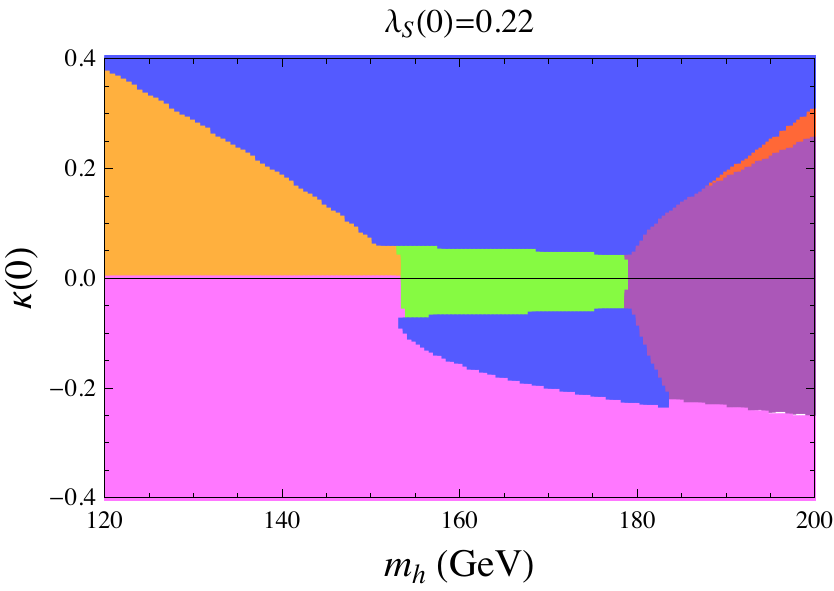} 
\end{array}$
\begin{center}
\includegraphics[scale=1.0]{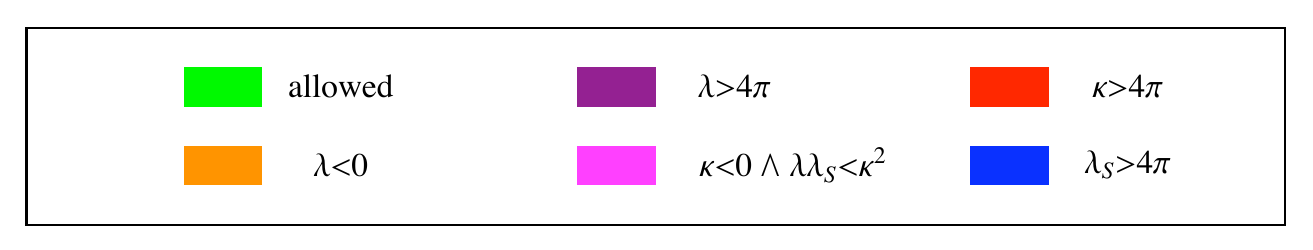} 
\end{center}
\caption{The vacuum stability and triviality constraints are applied to the parameter space (i.e. $\lambda (0)$, $\kappa (0)$, $\lambda_S (0)$) of the scalar sector of the minimal standard model in perturbation theory.  The non-minimal coupling to the gravitational field has been kept small with no suppression of the Higgs propagator, $s(t)=1.0$.  The constraints obtained apply up to $t=38$ which is slightly larger than the reduced Planck mass.  No allowed parameter space remains for $\lambda_S (0) \ge 0.23$.}
\label{nosuppression}
\end{figure*}

\begin{figure*}[ht]
$\begin{array}{cc}
\includegraphics[scale=1.0]{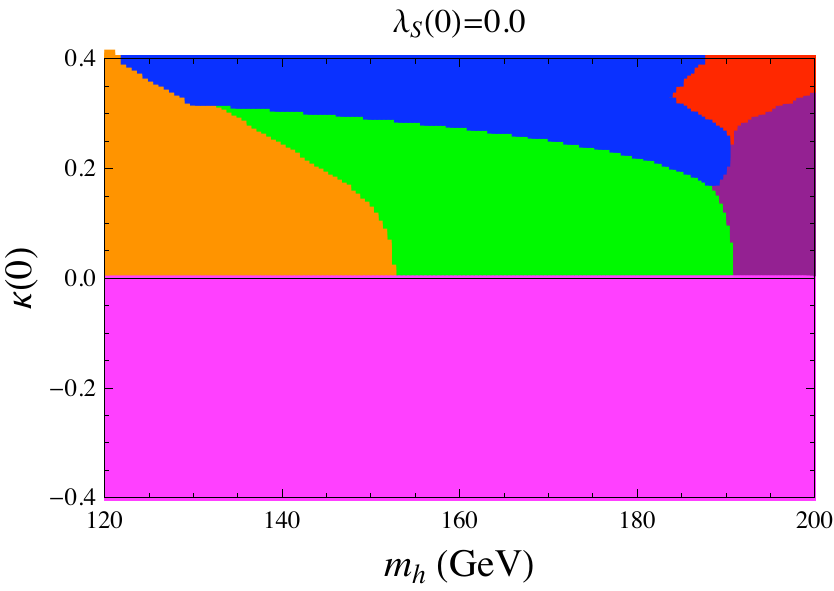} &
\includegraphics[scale=1.0]{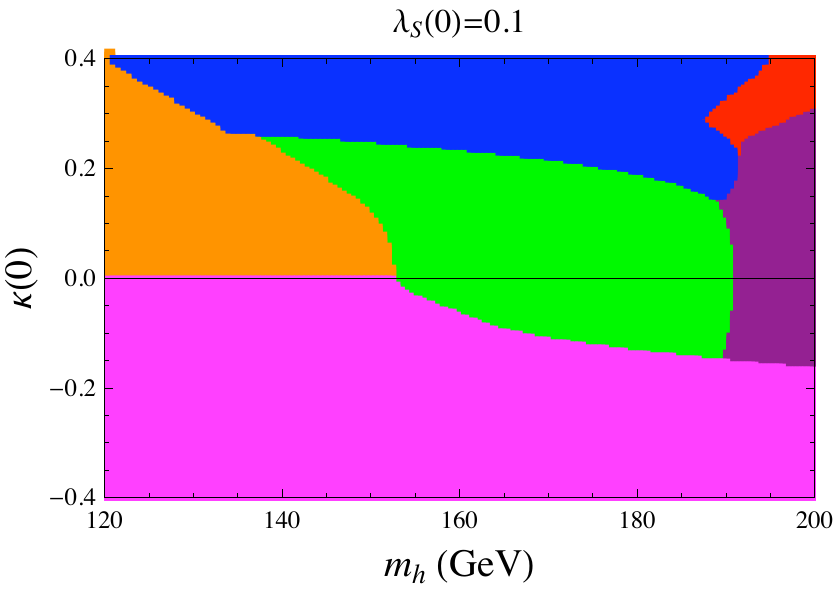} \\
\includegraphics[scale=1.0]{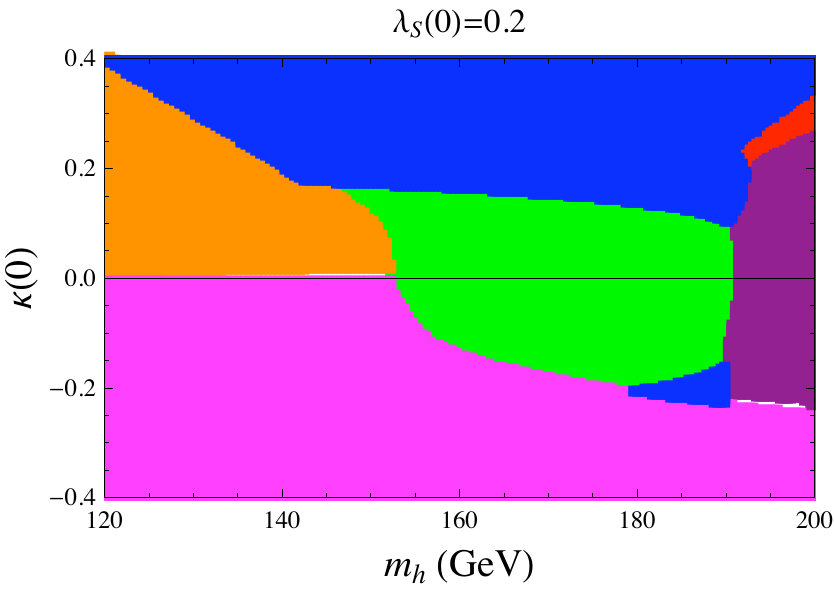} &
\includegraphics[scale=1.0]{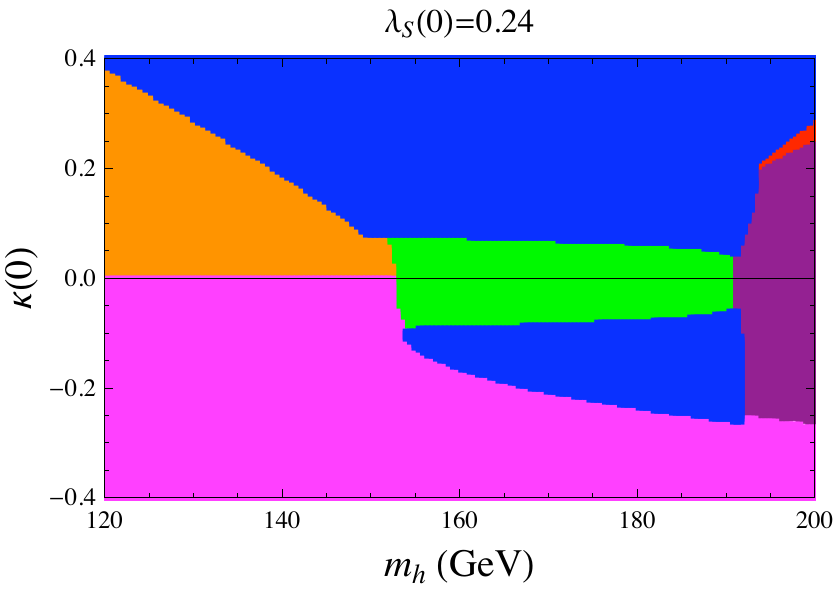} 
\end{array}$
\begin{center}
\includegraphics[scale=1.0]{legend.pdf} 
\end{center}
\caption{The vacuum stability and triviality constraints are applied to the parameter space of the scalar sector of the minimal standard model with large non-minimal coupling to the gravitational field, $\xi (0) = 10^4$.  The Higgs propagator is suppressed with the suppression factor given in Eq. (\ref{s(t)}) and the constraints apply to scales up to those typical of the onset of inflation, $t=34.5$.  No allowed parameter space remains for $\lambda_S (0) \ge 0.25$.}
\label{suppression}
\end{figure*}

In the following two sets of figures, the allowed parameter space, after the imposition of these vacuum stability and triviality constraints, is mapped out as a function of the various $t=0$ couplings ($m_h = \sqrt{2\lambda (0)}~ v$).  The triviality or vacuum stability bound that is violated for the various regions of the parameter space is indicated.  In Fig. \ref{nosuppression}, the parameter space constraints are applicable up to $t=38$ which corresponds to a scale slightly larger than the reduced Planck mass and with no suppression of the Higgs propagator ($s(t)=1$).  In the next panel, Fig. \ref{suppression}, the parameter space constraints are determined for the case where the Higgs propagator is suppressed and the vacuum stability and triviality bounds are imposed up to the typical scale for the onset of inflation, $t=34.5$.  Here the regions were determined with an initial value of the non-minimal gravitational coupling $\xi (0)=10^4$ and $\xi_S (0) =0$.  The size of the regions were found to be insensitive to the value of $\xi$ for a range of values $10^3 \le \xi (0) \le 10^4$.  The standard model without dark matter coupling corresponds to the $\kappa (0) = 0$ abscissa in the $\lambda_S (0)=0$ plot.  There are only minor differences in the allowed parameter space for the two different cases depicted in Figs. \ref{nosuppression} and \ref{suppression} which can be directly traced to either the non-propagation of the Higgs field for larger $t$ values or to the fact that the constraints are applied to a larger $t$ value in the unsuppressed Higgs propagator case. Thus, with a suppressed Higgs propagator, somewhat larger values,  $m_h \simeq 190 $ GeV, are allowed (as compared to $\sim $ 180 GeV with no suppression factor).

\newpage
\end{document}